# Second harmonic optical coherence tomography

Yi Jiang, Ivan Tomov, Yimin Wang, and Zhongping Chen*

*Beckman Laser Institute and Department of Biomedical Engineering, University of California, Irvine, Irvine, California 92612*

Second harmonic optical coherence tomography, which uses coherence gating of second-order nonlinear optical response of biological tissues for imaging, is described and demonstrated. Femtosecond laser pulses were used to excite second harmonic waves from collagen harvested from rat tail tendon and a reference nonlinear crystal. Second harmonic interference fringe signals were detected and used for image construction. Because of the strong dependence of second harmonic generation on molecular and tissue structures, this technique offers contrast and resolution enhancement to conventional optical coherence tomography.

Optical coherence tomography (OCT) is a noninvasive, noncontact imaging modality for cross-sectional imaging of biological tissue with micrometer scale resolution [1]. OCT uses coherence gating of backscattered light for tomographic imaging of tissue structures. Variations in the light scattering properties of tissue, due to inhomogeneities in the optical refractive index, provide imaging contrast. In many instances and especially in the early stages of disease, the change in tissue scattering properties between normal and diseased tissue is small and difficult to measure. To enhance the image contrast of OCT, a number of extensions have been developed: Optical Doppler Tomography combines the Doppler principle with OCT to obtain tissue structural image and blood flow simultaneously [2]; Spectroscopic OCT combines spectroscopic analysis with OCT to obtain the depth-resolved tissue absorption spectra [3]; Polarization Sensitive OCT combines polarization sensitive detection with OCT to determine tissue birefringence [4]. Engineered microsphere contrast agents for OCT have also been developed [5]. Recently, Coherent Anti-Stokes Raman Scattering interferometry has been proposed [6], but no image has been demonstrated to date using this method.

Optical second harmonic generation (SHG) is the lowest order nonlinear optical process where the second-order nonlinear optical susceptibility is responsible for the generation of light at the second harmonic (SH) frequency. Because the second-order nonlinear optical susceptibility is determined by detailed electronic configurations, molecular structures and symmetries, local morphologies, and ultrastructures, using SHG for biomedical imaging can give unique information regarding tissue structure symmetry [7]. The quadratic power dependence of SHG on the refractive index provides greater optical contrast for visualizing tissue structures than conventional linear reflectance microscopy does. SHG signal is typically detected in transmission mode for bulk transparent medium [8]. However, detection of SHG signal in reflection mode has been widely used to study nonlinear effects at surfaces and interfaces [7]. Recently, back-reflected SHG signals from unstained biological tissues has been investigated and used for imaging [9]. Quantitative second-harmonic generation microscopy in collagen has also been reported [10].

In this letter, we demonstrate an optical tomography technique, Second Harmonic Optical Coherence Tomography (SH-OCT), which combines the sample structural sensitivity of SHG with the coherence gating of OCT. The system consists of an interferometer illuminated by a low-coherence light source. If the sample possesses certain structures lacking a center of symmetry, the incoming light is converted into SH waves at the sample site. SH waves are also produced in the reference arm through a nonlinear crystal. The temporal interference pattern of these SH waves is then detected and used for image construction. Because of coherence gating, the depth resolution of imaging is determined by the coherence length of the SH wave and is independent of the objective. Thus, high depth resolution is achievable even using a low numerical aperture objective. SH-OCT also provides considerable imaging contrast and resolution enhancement over conventional OCT because of the high selectivity of SHG on tissue molecular structure and shorter measuring wavelength.

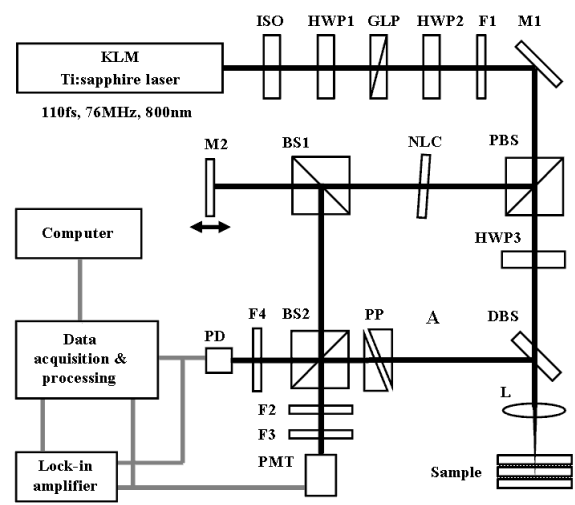

FIG. 1. SH-OCT experimental setup: ISO, isolator; HWP1-HWP3, half-wave plates; GLP, Glan laser polarizer; F1-F4, filters; M1-M2, mirrors; PBS, polarization beam splitter; NLC, nonlinear crystal BBO; BS1-BS2, broadband non-polarization beam splitter; DBS, dichroic beam splitter; L, objective; PP, prism pair dispersion compensator; PD, photodiode; PMT, photomultiplier tube.

The experimental configuration of SH-OCT system is shown in Fig. 1. The light source was a mode-locked femtosecond Ti:sapphire laser centered at 800 nm with 110 fs pulse duration and 76 MHz repetition rate. A long-wave pass filter (F1) filtered out the spurious SHG produced by the laser. A polarizing beam splitter (PBS) split the input beam into the two arms of the interferometer, and the split ratio was controlled by a half-wave plate (HWP2). In the reference arm, a 0.1 mm-thick nonlinear crystal (NLC) of ß-BaB2O4 (BBO) was oriented for type I phase matching to convert the fundamental wave into a SH wave at 400 nm. An important requirement for the nonlinear crystal in the sample arm is to be phase-matched for the whole spectrum of the fundamental radiation. A moving mirror (M2) acted as the delay line. In the signal arm, the fundamental wave was focused onto the sample by a low numerical aperture objective (L, N.A.=0.2, f=31.8 mm). When the sample had second-order nonlinear properties, the fundamental wave generated a SH wave. Back-reflected fundamental and SH waves were collimated by the same objective (L) and directed by a dichroic beam splitter (DBS) toward the combining beam splitter (BS2). The DBS reflected most of the SH wave and about 5% of the fundamental wave. The waves from both arms were recombined after passing BS2. By changing the optical path delay in the reference arm, the pulses overlapped temporally and interference fringes at SH and fundamental wavelengths were generated. With proper filtration provided by F2 and F3, the fundamental and SH interference fringe signals were detected by a photomultiplier tube (PMT) and a photodiode (PD) respectively. A prism-pair was inserted into the signal arm to compensate for the group-velocity dispersion of the fundamental and SH waves, thus enabling simultaneous detection of SH-OCT and conventional OCT signals.

To measure the coherence lengths in this hybrid OCT system, the sample was replaced by a polished GaAs crystal to produce surface SH waves (a nonlinear optical mirror) [7, 11]. Interferences at both wavelengths were measured as shown in Fig. 2, where SH interference occurs at exactly double the frequency of fundamental interference. It is well known that for a Gaussian pulse $I(t) \sim \exp[-(2t/t_p)^2 \ln 2]$ with pulse duration $t_p$, the coherence length $l_c$ is given by $l_c = (2\ln 2/p)(l_0^2/\Delta l)$, where $l_0$ is the center wavelength and $\Delta l = (2\ln 2/p)(l_0^2/ct_p)$ is the spectral width (full width at half-maximum). A relationship of $\Delta l_1/\Delta l_2 = 4/\sqrt{2}$ and $l_{c1}/l_{c2} = \sqrt{2}$ can be obtained from simple calculations, where $\Delta l_1$ and $\Delta l_2$ are the spectral widths of fundamental and SH waves, $l_{c1}$ and $l_{c2}$ are the coherence lengths of the fundamental and SH waves. The fundamental laser radiation has a spectral width of 8.1 nm (Fig. 3a). The spectral width of SH wave from the BBO crystal was measured to be 3.0 nm (Fig. 3b). The coherence lengths of the fundamental and SH waves in free space were measured to be 33 μm (Fig. 3c) and 24 μm (Fig.

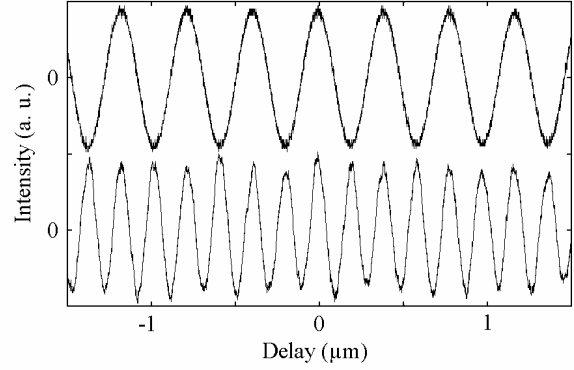

FIG. 2. Interference signals of fundamental waves (upper curve) and SH waves (lower curve). The SH interference signal is at double the frequency of the fundamental interference signal.

3d), respectively. All measured values are in good agreement with theoretic predictions.

The sample used in our study was Type I collagen harvested from rat tail tendon [8]. The sample consisted of two ~30 μm-thick collagen layers sandwiched among three 170 μm-thick glass slides (schematic shown in the top of Fig. 4). Using the Gaussian beam approximation, the objective (L) has a depth of focus of 0.52 mm. The average excitation power entering the objective was approximately 50 mW, and estimated peak power density at the beam waist in the sample was about 3.2 GW/cm². This peak intensity is two orders of magnitude smaller than the peak intensity threshold for the loss in cell viability demonstrated by König et al. [12]. The tomographic imaging experiment was conducted by scanning the mirror (M2) in the delay line and recording the fundamental and SH interference signals with a lock-in amplifier. The lock-in amplifier demodulated the interference fringe envelope signal with high sensitivity and precision. The measured OCT signals of one typical depth (z direction) scan are shown in Fig. 4. The conventional fundamental OCT signal in Fig. 4a shows the sandwich structure of the sample. The strong reflectance occurring at the first air-glass interface suppresses signals from the deeper layers. The SH-OCT signal in Fig. 4b shows two peaks that correspond to the two-layer structure since SH signals are only produced in the two collagen layers. Comparison of Fig. 4a and Fig. 4b shows that there is no SH-OCT signal coming from the air-glass interface, indicating that SH-OCT provides good contrast for nonlinear media. SH-OCT signals reveal information regarding the second-order nonlinear properties of the sample that cannot be provided by conventional OCT signals. Fig. 5 is a SH-OCT image of the center part of the sample, where two layers of collagen can be clearly observed.

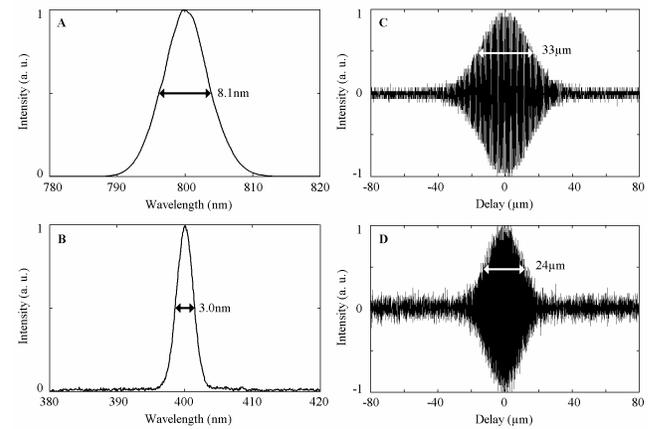

FIG. 3. Coherence length measurements of the hybrid OCT system: A, spectrum of the fundamental wave; B, spectrum of SH wave from the nonlinear crystal; C, Fundamental wave interference fringes; D, SH wave interference fringes.

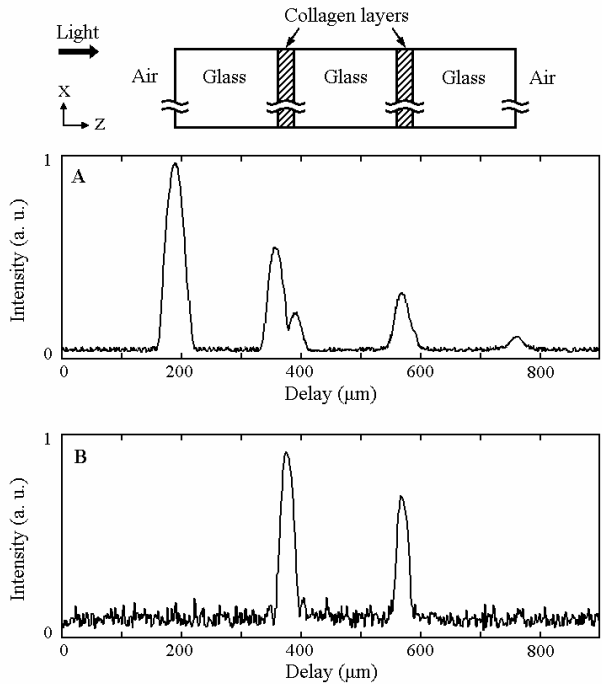

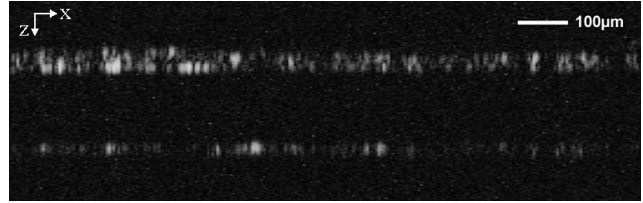

FIG. 5. A SH-OCT image of two collagen layers sandwiched among glass slides.

FIG. 4. SH-OCT and conventional OCT signals of one depth scan. The sample structure is shown in the top of this figure. A: Conventional OCT interference fringe envelope signal at the fundamental wavelength; B: SH-OCT interference fringe envelope signal.

SHG serves as a unique contrasting mechanism for tissue structure since SH signals are highly dependent on the orientation, birefringence, and local symmetry properties of the tissue. SHG efficiency in the collagen sample depends on orientation of collagen fibrils relative to the incident electrical field polarizations. In the experiment, the half-wave plate HWP3 was used to control the input beam polarization to the sample. Another half-wave plate optimized for the SH wavelength can be inserted into the reference arm after the reference crystal. By matching the SH wave polarization in both arms, collagen fibrils with different orientations can be preferentially highlighted to produce polarization dependent tomographic images.

In conclusion, we have presented a noninvasive optical tomography technique of second harmonic optical coherence tomography and experimentally demonstrated the feasibility of using this technique to image biological samples. SH-OCT offers enhanced image contrast and resolution to the conventional OCT using the same light source, which extends the capability of conventional OCT for detecting small changes in tissue and molecular structure and symmetry.

This work was supported by research grants awarded by the National Science Foundation (BES-86924) and National Institutes of Health (EB-00293, NCI-91717, RR-01192). Institute support from the Air Force Office of Scientific Research (F49620-00-1-0371) and the Beckman Laser Institute Endowment is also gratefully acknowledged.
* Electronic address: zchen@laser.bli.uci.edu